\documentclass[12pt]{iopart}
\usepackage{graphicx}
\usepackage{color}
\begin{document}

\title{Characterisation of longitudinal variation in photonic crystal fibre}

\author{Robert J. A. Francis-Jones, and Peter J. Mosley}

\address{Centre for Photonics and Photonic Materials, Department of Physics, University of Bath, Bath, BA2 7AY, United Kingdom}
\ead{r.j.a.francis-jones@bath.ac.uk}

\begin{abstract}
We present a method by which the degree of longitudinal variation in photonic crystal fibre (PCF) may be characterised through seeded four-wave mixing (FWM). Using an iterative numerical reconstruction, we created a model PCF that displays similar FWM phasematching properties across all measured length scales. Our results demonstrate that the structure of our PCF varies by less than $\pm 1\,\%$ and that the characteristic length of the variations is approximately 15\,cm.
\newline \\
{\bfseries OCIS}: (060.5295) Photonic crystal fibers; (060.2270) Fiber characterization; (190.4380) Nonlinear optics, four-wave mixing.
\end{abstract}



\section{Introduction}

This year marks the $20^{\textnormal{th}}$ anniversary of the first demonstration of light guidance in photonic crystal fibre (PCF) formed by an array of air holes in a solid silica glass matrix~\cite{Knight1996All_Silica_Single_Mode}. PCF has revolutionised the field of nonlinear fibre optics, enabling the development of light sources ranging from ultra-broad and bright supercontinuum~\cite{Wadsworth2002Supercontinuum_generation_in,Hooper2011Coherent-Supercontinuum-Generation} to sources of single photons~\cite{Sharping2004Quantum-correlated-twin-photons}. The success of PCF is primarily due to its capacity for engineering group-velocity dispersion (GVD) through changing the diameter, $d$, and pitch, $\Lambda$, of the air holes that form the cladding ~\cite{Knight2003Photonic-Crystal-Fibres}. This enables the transition between normal and anomalous GVD, known as the zero-dispersion wavelength (ZDW), to be tuned over a much wider range than in conventional fibre, yielding control over both linear propagation and nonlinear processes.

However, this flexibility also brings a high degree of sensitivity to fluctuations in the dimensions of the PCF structure. Some level of variation inevitably arises during the fabrication process, either due to imperfections in the fibre preform or small changes in the draw parameters. For some applications of PCF, such as supercontinuum generation pumped in the anomalous dispersion regime, these variations are insignificant and can be ignored. However for tasks that are more critically dependent on GVD, for example the fabrication of a PCF with all-normal dispersion or in applications of four-wave mixing (FWM), even small changes in the fibre structure can have highly detrimental effects.

As an example, we focus on the impact of structural variation on the use of spontaneous FWM in PCF for the generation of heralded single photons in pure quantum states\cite{Cohen2009Tailored-Photon-Pair-Generation, Halder2009Nonclassical-2-photon-interference, Soller2010Bridging-Visible-and-Telecom}. High-quality sources of single photons are a critical resource for developing photonic quantum technologies for information processing, communications, and metrology~\cite{OBrien2009Photonic-quantum-technologies}. Spontaneous FWM in fibre is a powerful technique for creating single photons using heralding: photon pairs created through FWM can be split up and one photon detected to herald the presence of its twin~\cite{McMillan2009Narrowband-High-Fidelity}. The dispersion-engineering capabilities of PCF give control over the properties of these photon pairs, critical to produce the high-purity states required for quantum-information processing \cite{Grice2001Eliminating-Frequency-and-Space-Time, Garay-Palmett2007Photon-Pair-State-Preparation, Cui2012Minimizing-the-Frequency-Correlation}. However, we show in this work, even small variations in PCF structure and dispersion can ruin these delicate quantum states.

To understand and mitigate the impact of such variation, we have used seeded FWM to probe the level of variation in PCF dispersion, followed by numerical reconstruction of likely dispersion profiles to relate our data to the structural parameters, $d$ and $\Lambda$. Our measurement is based on a process known as stimulated emission tomography (SET) conceived by Liscidini and Sipe to find efficiently the spectral probability distribution of photon pairs generated by parametric downconversion (PDC)~\cite{Liscidini2013Stimulated_Emission_Tomography}. A number of research groups have since used stimulated processes to determine the spectral properties of photon-pair sources based on both PDC~\cite{Eckstein2014High-resolution-spectral-characterization} and FWM ~\cite{Fang2014Fast_and_Highly_Resolved, Jizan2015Bi_Photon_Spectral, Fang2016Multidimensional-characterization-of-an-entangled}. When compared to methods of determining the two-photon probability distribution by photon counting~\cite{Mosley2008Heralded-Generation-of-Ultrafast,Avenhaus2009Fibre_Assisted_Single_Photon,Soller2010Bridging-Visible-and-Telecom}, SET-based techniques typically achieve higher resolution and are significantly faster. We describe how this enabled us to  measure rapidly a large number of PCF segments and hence determine the length scale of the structural variation.

\section{Four-wave mixing in uniform PCF}

The structure of a length of PCF dictates the wavelengths of light that can be generated by FWM because the process is parametric and requires perfect phasematching to operate efficiently. The wavelengths that can achieve phase matching are set by the dispersion, which is in turn controlled by the PCF structure. Hence FWM can be used to probe the structure of PCF.

We consider only FWM with a single (degenerate) pump pulse. As it propagates, the frequencies that can be generated, known as the signal and idler, are determined by energy and momentum conversation:
\begin{eqnarray}
	2\omega_{p} & = & \omega_{s} + \omega_{i}\label{eq:ECon},\\
	\Delta\beta & = & 2\beta_{p}(\omega_{p}) - \beta_{s}(\omega_{s}) - \beta_{i}(\omega_{i}) - 2\gamma P\label{eq:Phasemismatch}.
\end{eqnarray}
where $\omega_{j}$ and $\beta_{j}(\omega_j)$ are the frequency and propagation constant of pump, signal and idler, $j = p, s, i$, respectively and the final term on the right hand-side accounts for phase modulation by the pump pulse. We calculate the frequency-dependence of the propagation constants $\beta_{j}(\omega_j)$ from empirical relationships for standard PCF structures \cite{Saitoh2005Empirical-relations-for-simple} and assume that all fields propagate in the fundamental guided mode. Only pairs of signal and idler frequencies that satisfy Eq.~\ref{eq:ECon} and are phasematched ($\Delta\beta \approx 0$) will experience growth as the pump travels along the fibre. The ability to control the dispersion of the PCF through changes to the cladding structure provides a means by which specific signal and idler frequencies can be produced.

\begin{figure}
	\centering
	  \includegraphics[width = 0.99\textwidth]{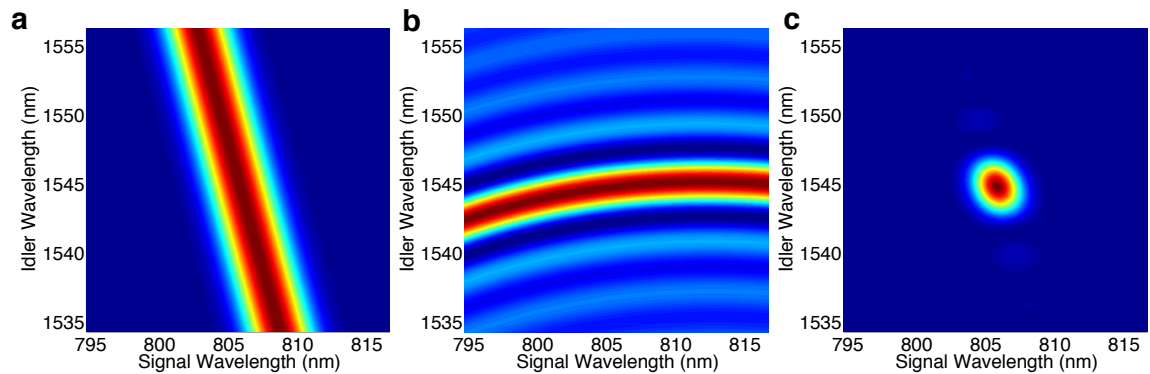}
    \caption{(a) Pump envelope function of an 800\,fs transform limited pulse with a Gaussian spectrum centered at 1064\,nm. (b) Real part of the phasematching function of a PCF with $\Lambda = 1.49 \mu m, d/\Lambda = 0.41$ designed to produce FWM at 805\,nm and 1545\,nm. (c) The resulting joint spectral intensity.}
    \label{fig:Homo_Sim_PMF_PEF_JSI}
\end{figure}

The relationship between the generated signal and idler frequencies can be visualised through the joint spectral intensity (JSI) distribution, $F(\omega_{s},\omega_{i})$. When FWM is used as a source of photon pairs, the JSI shows the probability of generating an idler photon with frequency $\omega_{i}$ given a signal photon with frequency $\omega_{s}$. The JSI is determined by the product of two functions: the pump envelope function $\alpha(\omega_{s} + \omega_{i})$ resulting from energy conservation and the phasematching function $\phi(\omega_{s},\omega_{i})$ that embodies momentum matching:
\begin{equation}
	F(\omega_{s},\omega_{i}) =  |f(\omega_{s},\omega_{i})|^{2} = |\alpha(\omega_{s} + \omega_{i})\cdot\phi(\omega_{s},\omega_{i})|^{2}.
    \label{eq:JSI_eq}
\end{equation}
When pumped by an ultrafast laser pulse, the bandwidth of the pump yields a range of potential solutions of Eq.~\ref{eq:ECon}. Therefore, $\alpha(\omega_{s}+\omega_{i})$ describes the variation in the signal and idler frequencies that are possible due to the finite bandwidth of the pump. For a pulse with a gaussian spectrum the pump envelope function is given by
\begin{equation}
	\alpha(\omega_{s} + \omega_{i}) = \exp{\left( - \frac{(\omega_{s} + \omega_{i} - 2\omega_{p0})^{2}}{4\sigma_{p}^2}\right)},
\end{equation}
where $\omega_{p0}$ is the central frequency of the pump and $\sigma_{p}$ is the $1/e$ bandwidth of the pump, as shown in Fig.~\ref{fig:Homo_Sim_PMF_PEF_JSI}a. We assume throughout that $\alpha(\omega_{s}+\omega_{i})$ remains constant along the length of the fibre, that is to say that the pump power is sufficiently low that the effects of self-phase modulation can be ignored, and that fibre attenuation and pump depletion are negligible.

The phasematching function, $\phi(\omega_{s},\omega_{i})$, shown in Fig.~\ref{fig:Homo_Sim_PMF_PEF_JSI}b describes the variations in the signal and idler frequencies that are possible given the phase mismatch that arises over a length $L$ of the PCF. For a homogeneous PCF, the propagation constants of the four fields do not vary longitudinally and $\phi(\omega_{s},\omega_{i})$ is given by the integral of the phase mismatch over the interaction length~\cite{Garay-Palmett2007Photon-Pair-State-Preparation}:
\begin{eqnarray}
	\phi(\omega_{s},\omega_{i}) & = & \chi^{(3)}\int_{0}^{L}e^{i\Delta\beta z}dz\label{eq:PMF_Int}.\\
								& = & 2\chi^{(3)} L\textnormal{sinc}\left(\frac{\Delta\beta L}{2}\right)\exp{\left(\frac{i\Delta\beta L}{2}\right)}. \label{eq:PMF_Homog}
\end{eqnarray}
A typical JSI for FWM in a homogeneous PCF is shown in Fig.~\ref{fig:Homo_Sim_PMF_PEF_JSI}c; it can be seen that the function has a single peak with small side-lobes as a result of the sinc function in $\phi(\omega_{s},\omega_{i})$. Note that, in this case, the frequencies of signal and idler are almost uncorrelated.

\section{The effect of structural variation on FWM}

\begin{figure}
	\centering
	\includegraphics[width = 0.8\textwidth]{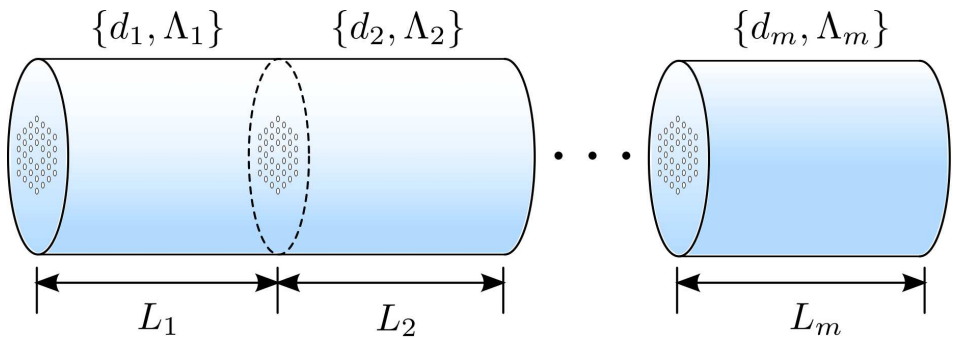}
	\caption{Schematic of a non-uniform PCF composed of $m$ homogeneous sections. Each of the $m$ sections is described by three parameters: $d$, $\Lambda$ and $L$.}
	\label{fig:Inhomo_PCF_Schem}
\end{figure}

Structural variations in PCF result in fluctuations in GVD and changes in the rate at which phase mismatch is accumulated along the fibre. The phasematching function of a non-uniform PCF can be modelled by dividing it into a series of $m$ segments each of which is homogeneous. Each segment is described by a set of parameters $\{d_{n},\Lambda_{n},L_{n}\}$ corresponding to the hole size, pitch and length of the $n$'th segment, giving rise to a phasemismatch of $\Delta \beta_{n}L_{n}$ with $\sum_{n}^{m} L_{n} = L$. The phasematching function of the complete inhomogeneous fibre can be found by integrating Eq.~\ref{eq:PMF_Int} in a piecewise manner resulting in \cite{Cui2012Spectral-Properties-of-Photon-Pairs}:
\begin{eqnarray}
	\phi(\omega_{s},\omega_{i}) &= L_{1}\textnormal{sinc}(\frac{\Delta \beta_{1}L_{1}}{2})\exp(i\frac{\Delta \beta_{1} L_{1}}{2}) \\
	& + \sum_{n = 2}^{m}  L_{n}\textnormal{sinc}(\frac{\Delta \beta_{n}L_{n}}{2})\exp(i\frac{\Delta \beta_{n} L_{n}}{2})\exp(i\sum_{l = 1}^{n-1}\Delta \beta_{l} L_{l}),
\label{eq:Inhomo_PMF}
\end{eqnarray}
corresponding to the coherent combination of the different phasematching functions of each homogeneous segment, with phase mismatch carried forward from the previous segments.

\begin{figure}
	\centering
    \includegraphics[width = 0.95\textwidth]{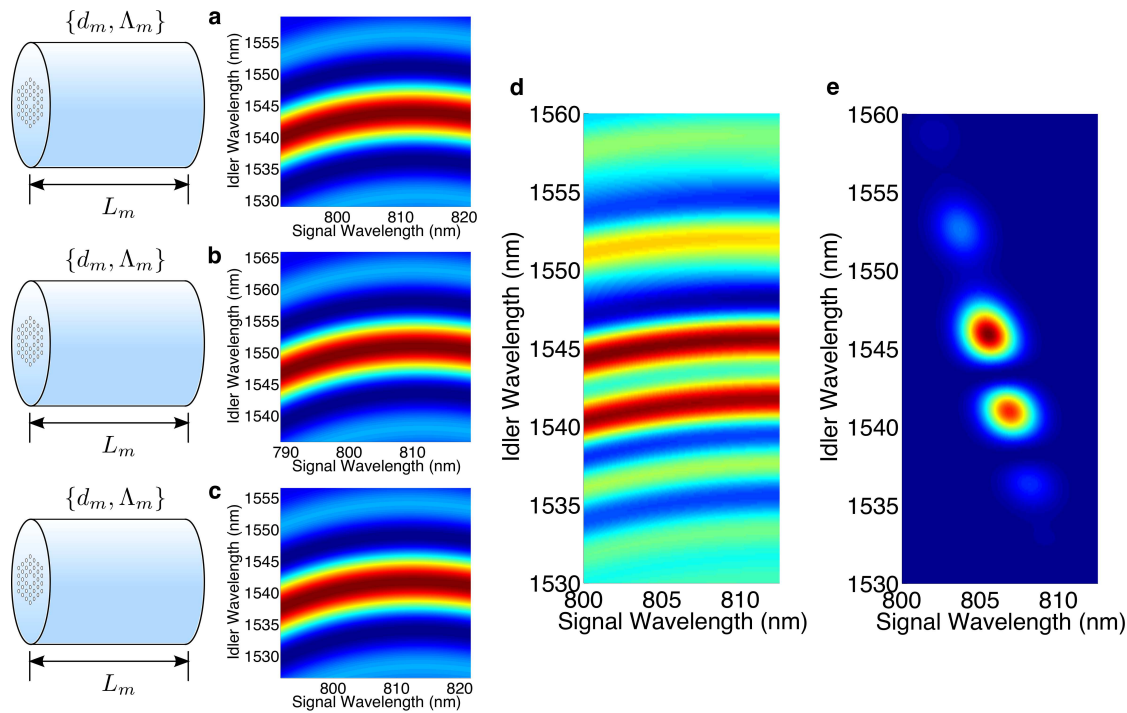}
    \caption{(a - c) Phasematching functions of three different sections of homogeneous PCF each 0.083m in length and with nominal values of $\Lambda = 1.49 \mu m, d/\Lambda = 0.41$. (d) PMF of inhomogeneous fibre formed by coherent addition of phasematching functions (a)--(c). (e) Resulting joint spectral intensity for inhomogeneous PCF.}
	\label{fig:Inhomo_Sims}
\end{figure}

In Fig.~\ref{fig:Inhomo_Sims} we show how the JSI of an inhomogeneous PCF 0.25\,m in length can be constructed from three homogeneous segments with structural parameters drawn from a normal distribution with a variance of $1\%$. Each section has a different phasematching function given by Eq.~\ref{eq:PMF_Homog} that combine to yield the complete phasematching function shown in panel d. When multiplied by the pump function shown in Fig.~\ref{fig:Homo_Sim_PMF_PEF_JSI} a, we obtain the JSI for the inhomogeneous PCF displayed in Fig.~\ref{fig:Inhomo_Sims}. It is clear to see that the changes in structure and therefore dispersion along the length of the PCF have drastically altered the FWM JSI compared to that of the homogeneous PCF in Fig.~\ref{fig:Homo_Sim_PMF_PEF_JSI} c. Whereas the JSI for the homogeneous fibre exhibited a single area of high probability, for the inhomogeneous PCF additional high probability features appear within the JSI that increase the spectral correlation between signal and idler.

In high-intensity FWM, these additional features in the JSI increase the bandwidth of the generated beams and reduce spectral brightness. However, when FWM is used for photon-pair generation, their effect is far more detrimental. In order to generate heralded single photons in pure quantum states, the JSI must be engineered to minimise frequency correlation between signal and idler. This is enabled by dispersion control in PCF, as shown for the homogeneous case in Fig.~\ref{fig:Homo_Sim_PMF_PEF_JSI} c, but  additional features in the JSI caused by variations in the PCF structure introduce frequency correlation and drastically reduce the purity of the resulting heralded photons making them unsuitable for quantum information applications. Hence to generate pure single photons it is critical to characterise the level of variation accurately, and understand the length scale over which the structure fluctuates.
 
\section{Measurements of FWM joint spectra}

\begin{figure}
	\centering
    	\includegraphics[width = 0.8\textwidth]{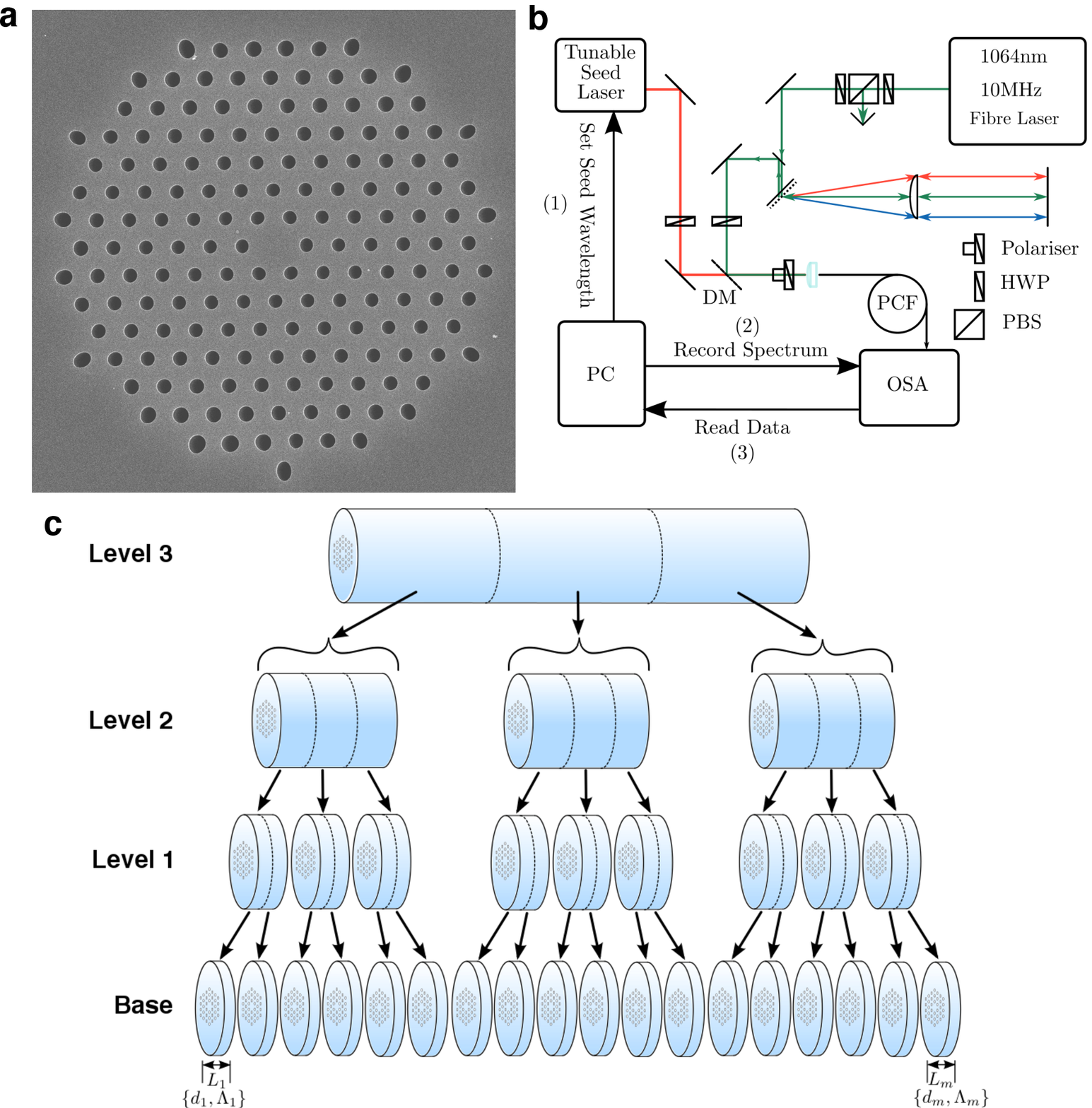}
	\caption{(a) Scanning electron micrograph of cleaved PCF end face with a pitch of $d = 1.49 \mu m$. (b) Schematic of stimulated emission tomography measurement; see text for details. (c) Schematic of the measurement tree. The complete PCF consists of a single 3\,m length (level 3). This was divided into three 1\,m lengths (level 2), each subsequently split into a further three 30\,cm lengths (level 1). Finally, each of the level 1 sections was cut into two 15\,cm lengths (base segments).}
	\label{fig:SET_Set_Up}
\end{figure}

An 8-ring PCF was fabricated using the stack-and-draw technique. The PCF was designed to generate signal and idler wavelengths in the region of 800\,nm and 1550\,nm respectively when pumped at 1064\,nm. A scanning electron micrograph of the end face is shown in Fig.~\ref{fig:SET_Set_Up}a. One capillary in the outermost ring of the cladding was replaced by a solid silica rod to create a missing hole used to orientate the PCF. The FWM JSIs of various lengths of the PCF were mapped using seeded FWM to emulate the SET technique~\cite{Liscidini2013Stimulated_Emission_Tomography}. It has been shown for FWM processes that when a seed field is injected at the idler wavelength the output at the conjugate signal frequency is proportional to the JSI~\cite{Fang2014Fast_and_Highly_Resolved}. Hence by tuning a narrow-band seed field across a range of idler wavelengths the JSI can be recovered from a series of measurements of the signal spectrum.

A schematic of the experimental set-up is shown in Fig~\ref{fig:SET_Set_Up}b. A 1064~nm 10~MHz fibre laser (\emph{Fianium} FemtoPower-PP) was used to pump the FWM. The beam from this laser first passed through a half-wave plate and polarising beam splitter for power control before entering a 4f-spectrometer in which the central wavelength and bandwidth of the pump pulses were adjusted. A further half-wave plate and polariser were used to set the polarisation in the PCF. For the seed field a tuneable CW laser (\emph{ThorLabs INTUN-1550}) with a linewidth of approximately 125~kHz was used at the idler wavelength. The co-polarised seed and pump fields were mixed on a dichroic mirror and coupled into the PCF using and aspheric lens. Each fibre segment under measurement was placed in a bare fibre adapter with the missing hole marker aligned with the pump and seed polarisation. The output of the PCF was coupled directly to an optical spectral analyser (OSA).

A computer controlled both the tunable seed laser and the OSA to allow fast data acquisition at high resolution. First the wavelength of the seed laser was set, following this the OSA scan was triggered. Once completed, the recorded stimulated signal spectrum was read out over GPIB to the PC and a new seed wavelength selected. The process was repeated for 100 different seed wavelengths. The JSI was recovered by stacking each stimulated signal spectrum into a 2D plot.

We began with a 3\,m length of the PCF and recorded its JSI. This 3\,m sample was cut into lengths of 1\,m and the JSI of each measured. Subsequently each 1\,m length was cut into 30\,cm sections, and each of 30\,cm section cut in to two 15\,cm pieces, with the JSI of each section measured at each stage. The overall measurement tree is shown in Fig.~\ref{fig:SET_Set_Up}c.

\begin{figure*}
	\centering
	\includegraphics[width = \textwidth]{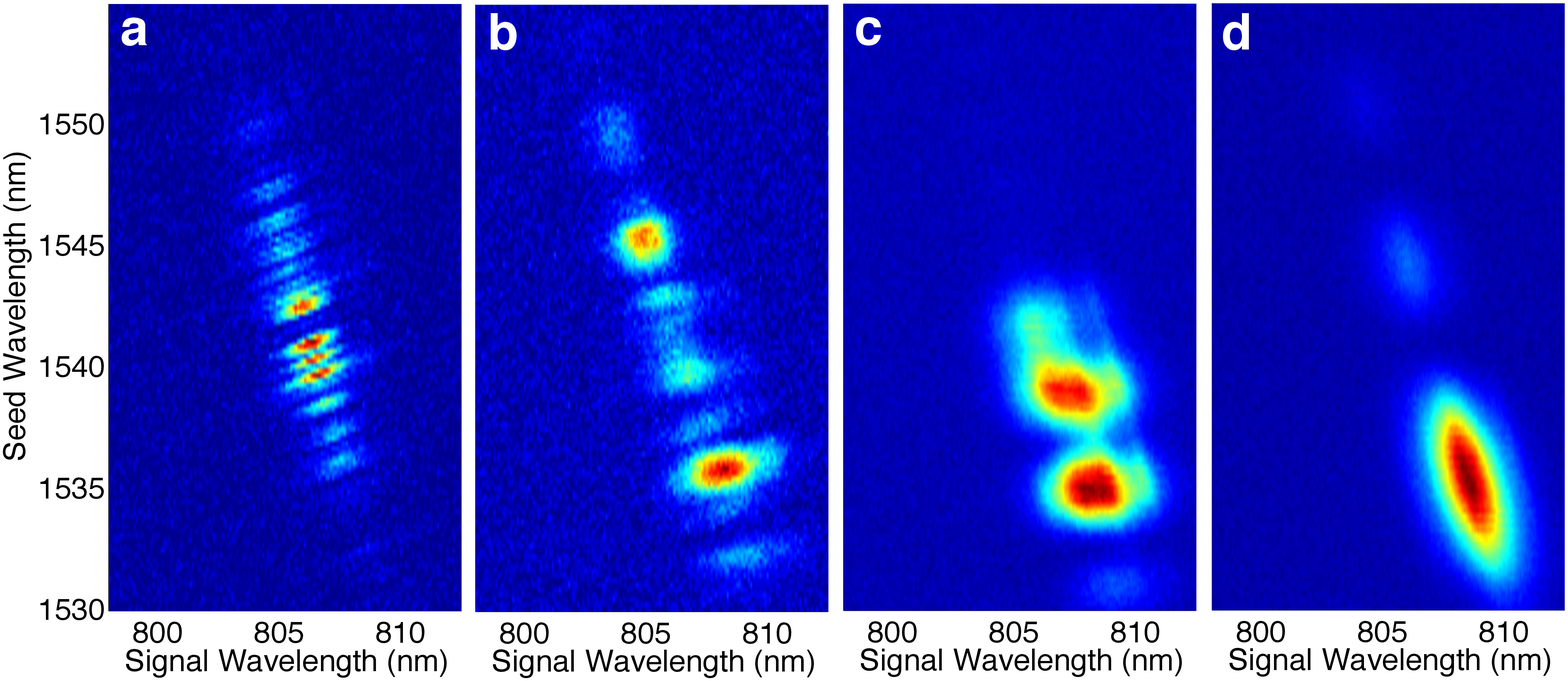}
	\caption{A selection of JSIs measured for different fibre lengths. (a) Level 1 segment with L = 3\,m. (b) Level 2 segment with L = 1\,m. (c) Level 3 segment taken from (b) with L = 30\,cm. (d) Base segment taken from (c) with L = 15\,cm.}
	\label{fig:Stim_JSIs}
\end{figure*}

A selection of the recorded JSIs at the different length scales are shown in Fig.~\ref{fig:Stim_JSIs}. We see that the level of inhomogeneity decreases as the fibre is cut up into shorter pieces, shown by the smaller number of high-intensity features in the JSIs of shorter lengths, until finally in Fig.\ref{fig:Stim_JSIs}d the sample is approximately homogeneous with a single high-intensity lobe. For the 15\,cm segments, the bandwidth of the phasematching function was now so large compared to that of the pump that the the JSI was dominated by the pump envelope function. In addition, a clear trend between sub-segments from the same parent segment was seen. Fig.~\ref{fig:Stim_JSIs} shows JSIs recorded for sub-segments taken from a single parent segment; the dominant features in the parent can be seen in the sub-segments as one would expect.

\section{Numerical reconstruction of PCF structure}

We have developed a numerical reconstruction technique to reconstruct the JSI of each parent segment using the JSIs of its sub-segments with the appropriate phase applied. We used the set of 15\,cm lengths that are approximately homogeneous, which we term the base segments, as a starting point to estimate numerically the structural parameters present in the original fibre.

To begin, the uniform FWM phasematching function in Eq.~\ref{eq:PMF_Int} was used to reconstruct the JSI of each of the base segments. Starting from nominal values for hole diameter $d$ and pitch $\Lambda$, the resulting phasematching function was combined with a standard pump envelope function that matched the experimental conditions. The parameters $d$ and $\Lambda$ were then iterated to fit a reconstructed JSI for each base segment to the measured data. The length of the fibre in the model was taken from  measurements of the fibre segments under test.
\begin{figure*}
	\centering
	\includegraphics[width = 0.7\textwidth]{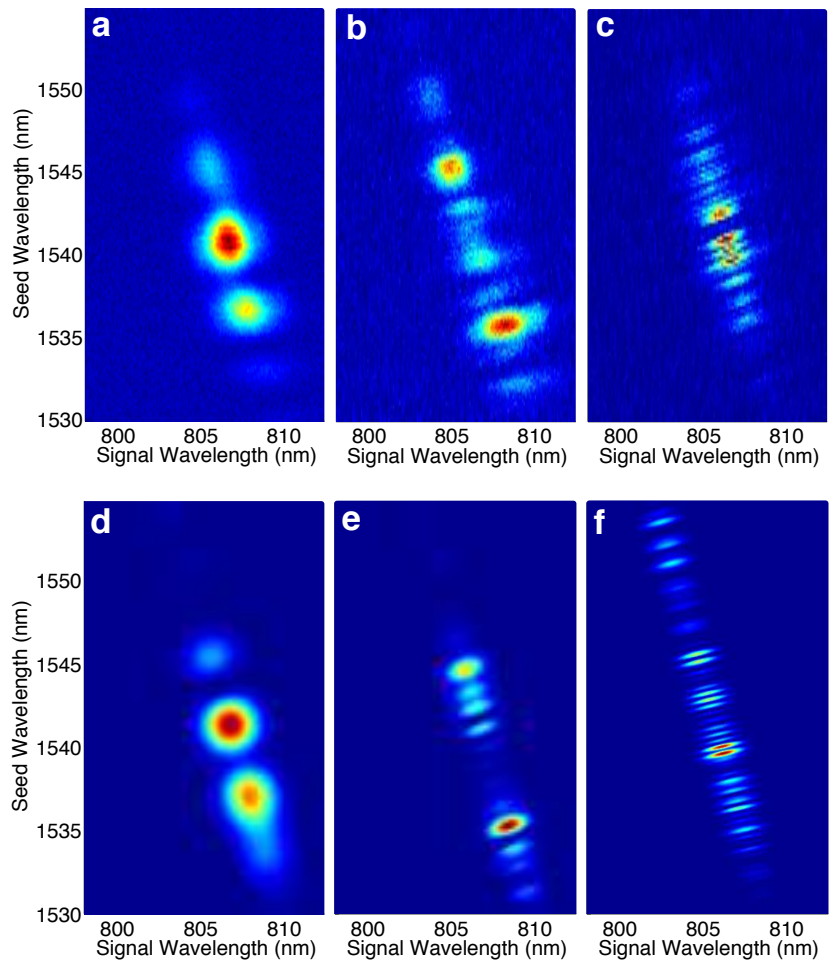}
	\caption{A comparison of measured JSIs (a)--(c) with the corresponding reconstructed JSIs (d)--(f). (a, d) Level 3 segment with L = 30 cm. (b, e) Level 2 segment with L = 1m. (c, f) Level 1 segment L = 3m.}
	\label{fig:Recon_JSIs}
\end{figure*}

The set of fitted structural parameters of the base segments formed the basis for the reconstruction of the JSIs of the longer segments.  First, the level 1 segments composed of two base segments were reconstructed by  combining the phasematching functions of the relevant base segments  using Eq.~\ref{eq:Inhomo_PMF}. The experimentally measured JSI and the numerically reconstructed JSI were then compared to evaluate the quality of the fitted structural parameters. To improve the quality of the fit, the fitting parameters of the base segments were adjusted and the JSI recalculated. This new JSI was compared to the measured JSI; if the new JSI was a better fit to the data the new set of fitted parameters were accepted, if not the changes were rejected. The process was repeated to optimise iteratively the overlap between measurements and simulation. Once this was achieved for all level 1 segments, these were combined as shown in Fig.~\ref{fig:SET_Set_Up}c to form the level 2 segments. The same optimisation procedure was applied, varying the parameters of the base segments to correct the fit. Finally, the full 3\,m fibre (level 3) consisting of the 18 base segments was reconstructed by combining the three level 2 segments together in the same manner. A comparison of the measured and reconstructed JSIs for the different stages of the reconstruction is shown in Fig.~\ref{fig:Recon_JSIs}.

\begin{figure}
	\centering
	\includegraphics[width = 0.6\textwidth]{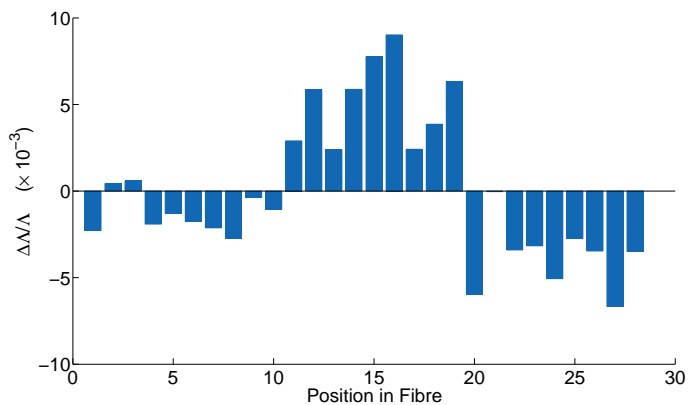}
	\caption{Fractional variation of the reconstructed pitch as a function of position along the fibre.}
	\label{fig:Fitting_Distribution}
\end{figure}

To simplify the reconstruction it was assumed that the majority of the inhomogeneity was introduced through fluctuations in the pitch, $\Lambda$, along the fibre. This allowed the reconstruction to be carried out using fewer parameters whilst still maintaining a good fit to the experimental data. For each base segment the hole size $d$ was fixed once the initial parameters were found. From the complete reconstructed PCF the fluctuation in the pitch along the fibre can be visualised as the fractional variance in the pitch about the mean of the entire fibre as shown in Fig.~\ref{fig:Fitting_Distribution}. 
The fractional variance in the pitch about the mean for the final parameter set shows variation over two different length scales. Firstly, on the longest length scales there is a large variation roughly every metre. Secondly, within each of those metre lengths, variations are still present but the difference is much smaller. On average, the pitch of all of the sub-segments lie within $\pm$1\% of the mean, in agreement with the 1\% variation in the outer diameter of the fibre that was achieved during fabrication.

\section{Conclusion}

In conclusion, we have demonstrated a technique for determining the degree of structural variation in PCF by measuring seeded FWM and carrying out a numerical reconstruction. Hence we infer the level of variation in our PCF to be below $\pm 1\%$ and the length scale of the variations to be around 15\,cm. Not only can these methods be used to determine the suitability of PCF samples for use in high-purity heralded single photon sources, where any inhomogeneity in the fibre introduces additional unwanted spectral correlations in the two-photon state, but also in other uses of PCF, for example in bright FWM sources that require high spectral brightness. Our work demonstrates that these measurements can be made using widely available equipment over short time scales and that a good estimate of the level of variation can be made without destroying the fibre.

\section*{Funding Information}
This work was funded by the UK EPSRC First Grant scheme (EP/K022407/1) and the EPSRC Quantum Technology Hub \textit{Networked Quantum Information Technologies} (EP/M013243/1).

\end{document}